\documentclass[aps,psfig,epsfig,preprint]
{revtex4}
\usepackage{amssymb}
\usepackage{epsfig}
\usepackage{color}
\usepackage{amsmath, amsthm}
\usepackage{amsmath, amsfonts}
\usepackage{amsmath}
\usepackage[all]{xy}

\makeatletter

\newcommand{\Rmnum}[1]{\expandafter\@slowromancap\romannumeral #1@}
\makeatother

\newcommand{\ep}{\epsilon}

\newcommand{\pa}{\partial}

\begin{document}
\draft
\title{
{\normalsize \hskip4.2in USTC-ICTS-05-15} \\{\bf Hamiltonian
Formalism of the de-Sitter Invariant Special Relativity  }}
\author{ Mu-Lin Yan\footnote{e-mail address: mlyan@ustc.edu.cn; corresponding
author.}, Neng-Chao Xiao, Wei Huang}

\affiliation{}

\address{
 Interdisciplinary Center for Theoretical Study, University
of Science and Technology of China, \\ Hefei, Anhui 230026, China}

\author{ Si Li}
\address{
 Department of Mathematics, University
of Science and Technology of China, \\ Hefei, Anhui 230026, China}

%\begin{document}
%\title{{\normalsize \hskip4.2in USTC-ICTS-05-03} \\{\bf Proton-Antiproton Annihilation in Baryonium }}
%\author{Gui-Jun Ding, Mu-Lin Yan\thanks{e-mail address: mlyan@ustc.edu.cn; corresponding author.} \\
%\small{Interdisciplinary Center for Theoretical Study} \\
% \small {University
%of Science and Technology of China, Hefei, Anhui 230026, China}}
\date{}
%\maketitle
\begin{abstract}
\noindent Lagrangian of the Einstein's special relativity with
universal parameter $c$ ($\mathcal{SR}_c$)  is invariant under
Poincar\'e transformation which preserves Lorentz metric
$\eta_{\mu\nu}$. The $\mathcal{SR}_c$  has been extended to be one
which is invariant under de Sitter transformation that preserves so
called Beltrami metric $B_{\mu\nu}$. There are two universal
parameters $c$ and $R$ in this Special Relativity (denote it as
$\mathcal{SR}_{cR}$). The Lagrangian-Hamiltonian formulism of
$\mathcal{SR}_{cR}$ is formulated  in this paper. The canonic
energy, canonic momenta, and 10 Noether charges corresponding to the
space-time's de Sitter symmetry are derived.  The canonical
quantization of the mechanics for $\mathcal{SR}_{cR}$-free particle
is performed. The physics related to it is discussed.

\vskip0.2in


\end{abstract}

\maketitle

\section{Introduction}
\noindent { Einstein's Special Relativity is the cornerstone of
physics. The theory indicates the space-time metric is
$\eta_{\mu\nu}=diag\{+,-,-,-\}$. The most general transformation to
preserve metric $\eta_{\mu\nu}$ is Poincar\'e group (or
inhomogeneous Lorentz group $ISO(1,3)$). It is well known that the
Poincar\'e group is the limit of the de Sitter group with  sphere
radius $R\rightarrow \infty$. Thus a natural question arisen from
this fact is whether there exists or not another type of de Sitter
transformation with $R \rightarrow finite$ which also leads to a
special relativity theory. In 1970's, K.H.Look (Qi-Keng Lu) and his
collaborators have pursued this question and got positive
answer\cite{look}\cite{Lu74}. In the recent years, some interesting
studies on Lu's theory in \cite{Guo1}\cite{Guo2} are stimulated by
the recent observations which show that there should be a positive
cosmological constant\cite{Ben}\cite{Teg}. In
\cite{Guo1}\cite{Guo2}, the length parameter $R$ in Lu's theory has
been identified as $\sqrt{3/\Lambda}$ where $\Lambda$ is the
cosmological constant. In this present paper, we try to study and to
reexamine   Lu's theory in Lagrangian-Hamiltonian formalism. Lu's
theory will be called as the de-Sitter Invariant Special Relativity
hereafter. }

Inertial motion law for free particles is the foundation of
mechanics. This law states that in the inertial reference frames the
free particle (i.e., without any  force acting on it) will move
along straight line and with constant coordinate velocities. The
Newtonian mechanics is the first mechanical theory built on this
foundation and without  any universal parameters. The Lagrangian for
free particle is:
\begin{equation}\label{1}
L_{\rm Newton}={1\over 2}m_0 {v}^2,
\end{equation}
where $m_0$ is the mass of the particle,  ${\bf v}=\dot{\bf x}$ is
the velocity and $v^2={\bf v}^2$. We may regard it as a
parameter-free
 realization of the inertial motion law. The second
mechanic theory realizing this inertial motion law is the Einstein's
Special Relativity with {\it one universal parameter $c$} (the
velocity of light). Denoting it as $\mathcal{SR}_c$, the Lagrangian
of free particle is (e.g., see, \cite{Landau}):
\begin{equation}\label{2}
 L_{c}=-m_0c{ds\over dt}=-m_0c{\sqrt{\eta_{\mu\nu}dx^\mu dx^\nu}\over dt}
 =-m_0c^2 \sqrt{1+{\eta_{ij}\dot{x}^i\dot{x}^j\over
 c^2}}=  -m_0c^2 \sqrt{1-{v^2\over c^2}},
\end{equation}
where Lorentz metric $\eta_{\mu\nu}={\rm diag}\{+,-,-,-\}$,
$dx^\mu=\{d(ct), dx^1, dx^2, dx^3 \}$ and $i,j=1, 2, 3.$ By means
of the Lagrange-Hamilton mechanics formulation, the particle's
momentum and Hamiltonian read
\begin{eqnarray}\label{3}
p_i&=&{\pa L_c \over \pa \dot{x}^i}={-m_0\dot{x}^j\eta_{ij} \over
\sqrt{1-{v^2\over c^2}}}, \\
\label{4} H&=& p_i \dot{x}^i-L_c=c\sqrt{-\eta^{ij}p_ip_j+m_0^2c^2}.
\end{eqnarray}
It is easy to check that when $c\rightarrow \infty$ the Special
Relativity back to the Newtonian mechanics, i.e.,
\begin{equation}\label{5}
L_{c}|_{c\rightarrow \infty}=L_{\rm Newton}+ {\rm constant}.
\end{equation}

An interesting and challenging question is  whether a mechanical
realization of the inertial motion law with {\it two universal
parameters} can be formulated or not. Surprisingly, the answer to
it is confirmative, and actually such a theory has already existed
in literature even though it is still less known so far.  About
thirty five years ago, K.H.Look (Qi-Keng Lu) found out that the
velocity of motion of the free particle along the geodesic line in
the de Sitter({\it dS})-space with Beltrami metric is constant,
and the geodesic is straight line\cite{look} \cite{Lu74}. This
theory is just the de Sitter invariant special relativity
mentioned  above. In Lu's theory, there are two universal
parameters: the light velocity $c$ and the de Sitter sphere radius
$R$ (or original notation $\lambda=1/R^2 $ used in \cite{look} and
\cite{Lu74}). The coordinate-transformation to preserve the
Beltrami metric has also been derived in \cite{look} \cite{Lu74}.
This means that the realization of the inertial motion law with
{\it two universal parameters} has been formulated. The theory
will be shortly denoted as $\mathcal{SR}_{cR}$ due to the
existence of two universal parameters $c$ and $R$ in the theory.
In this present paper, we try to provide a Lagrangian-Hamiltonian
formulation to illustrate the free-particle mechanics in the
de-Sitter invariant special relativity.

It is well known that the Lagrangian-Hamiltonian formulation in the
mechanics theory provides a sound foundation to discuss the
particle's motion, to deduce the particle's canonical (or conjugate)
momenta and the canonical energy (or Hamiltonian), { to derive the
Noether's charges corresponding to the symmetries,  and to over the
classical mechanics for constructing the quantum mechanics,} and so
on. In the previous works on
$\mathcal{SR}_{cR}$\cite{look,Lu74,Guo1,Guo2}, the
free-particle-motion in the space-time with Beltrami metric was
discussed by means of solving the geodesic equation, and it has been
found that the velocity of the particle is a constant. { This
remarkable claim should be reconfirmed in Lagrangian-Hamiltonian
formulation. Especially, because any reliable quantization
procedures of a classical mechanics theory  rely upon the theory's
Lagrangian-Hamiltonian formulation,  it is a basic task to determine
 the system's canonical momenta and the Hamiltonian.
To $\mathcal{SR}_{cR}$, the particle's canonical momenta and
Hamiltonian are unusual and somewhat subtle, which have to be
derived. Furthermore, the Noether's charges in $\mathcal{SR}_{cR}$,
which are the quantities in physics, should also be derived in this
formulation. For all these purposes, a systematic and careful study
on the Lagrangian-Hamiltonian formulation for $\mathcal{SR}_{cR}$ is
necessary.}

{Eq.(\ref{2}) shows the Lagrangian of free particle in
$\mathcal{SR}_c$ (i.e., ordinary Einstein's special relativity) is
time- and coordinate-independent (or $x^i$ are cyclic coordinates)
. So, both Hamiltonian and canonical momenta are motion of
constants. Furthermore, the most general space-time transformation
preserving $\eta_{\mu\nu}$ in $\mathcal{SR}_c$ is simply the
Poincar\'{e} transformation group (or inhomogeneous Lorentz group)
{\it ISO(1,3)}. Therefore conserved Noether charges are just its
Hamiltonian, canonic momenta, the angular momenta (and plus 3
Lorentz boost charges). All of these are well known. To
$\mathcal{SR}_{cR}$, however, the situation is much more
complicated than in $\mathcal{SR}_c$. Because the Beltrami metric
is time- and coordinate-dependent, we face a mechanical system
with time-dependent Hamiltonian and without any cyclic
coordinates. The space-time transformation preserving Beltrami
metric is a sort of de-Sitter transformation. In this case,
 a careful enough revisiting to the classical mechanics with time- and coordinate-dependent
Lagrangian is necessary for getting convincible conclusions. It
will be found out that    the Hamiltonian (or canonical energy),
canonical momenta are different from the conserved Noether charges
corresponding the external space-time symmetry of
$\mathcal{SR}_{cR}$. The latter are energy and momenta in physics,
and the former are the canonical quantities which are also useful
for mechanics, especially for the quantization of the system. }

Following $\mathcal{SR}_c$, in the framework of $\mathcal{SR}_{cR}$,
the wave equation of relativistical quantum mechanics is derived in
this paper by means of the standard canonic quantization procedure:
1) The Hamiltonian  mechanics leads to quantum canonic equations,
then Hamiltonian operator $\hat{H}$ and canonical momentum operators
$\hat{\pi}_i$ are defined; 2) By the mechanics again, the dispersion
relation between $\hat{H}$ and $\hat{\pi}_i$ is obtained, and hence
the we achieve the wave equation for the $\mathcal{SR}_{cR}$ quantum
mechanics. Due to existence of $x^i\hat{\pi}_i$-terms in the
time-dependent Hamiltonian $\hat{H}$, the operator ordering has to
be taken care. In our quantization scheme a generalized Weyl
ordering is taken, in which the external space-time symmetry of
$\mathcal{SR}_{cR}$ is preserved. This indicates that
$\mathcal{SR}_{cR}$ is consistent with the principle of quantum
mechanics.

{The contents of this paper are organized as follows: In section II,
we show explicitly that the Euler-Lagrangian equations are
equivalent to the geodesic equations for generic metric
$g_{\mu\nu}$. In section III, we construct the Lagrangian for
$\mathcal{SR}_{cR}$ by means of the Beltrami metrics, and solve the
equation of motions of free particle. The Hamiltonian and the
canonical momenta are also derived. Section IV is devoted to
calculate the Noether charges corresponding to external space-time
symmetry of $\mathcal{SR}_{cR}$. In Section V we discuss the
quantization of the system. Finally, we summarize our results
briefly. In the Appendix, we show how to derive space-time
transformation to preserve Beltrami metric following
refs.\cite{look}\cite{Lu74}. }

\section{Equation of motion for free particle in the space-time with
metric ~  $ g_{\mu\nu}(x)$}

The motion of a free material particle is determined in the special
theories of relativity from the principle of least action,
\begin{equation}\label{2.1}
\delta S \equiv \delta \int L(t, x^i, \dot{x}^i)dt= -m_0c\delta \int
ds =0,
\end{equation}
where $S$ is the action integral,  $ds=\sqrt{ g_{\mu \nu}dx^\mu
dx^\nu}$ is the the space-time interval. $g_{\mu\nu}=\eta_{\mu\nu}$
for $\mathcal{SR}_c$, but for $\mathcal{SR}_{cR}$, $g_{\mu\nu}$
should be Beltrami metric. Generally, from eq.(\ref{2.1}), we have
\begin{equation}\label{2.2}
 {L(t,x^\mu, \dot{x}^\mu)}=
 -m_0c {ds \over dt}=-m_0c\sqrt{g_{\mu \nu}{dx^\mu \over dt}{dx^\nu \over
 dt}},
\end{equation}
Variating the action with respect to $x^\mu$ we get 4-dimensional
Euler-Lagrangian equation where variables $x^0$ and $\dot{x}^0$
emerge as independent variables:
\begin{eqnarray}\label{2.3}
{\pa L \over \pa x^\lambda}-{d \over d t}{\pa L \over
\pa{\dot{x}^\lambda}}=0
\end{eqnarray}
where $t$ serves as a parameter rather than the physical coordinate
time, $\dot{x}^\lambda={{d x^\lambda} \over dt}$ and $\lambda$ runs
over all the space-time indices including $\lambda=0$. Obviously,
they are equations of motion, but not the standard Euler-Lagrangian
equations in the Hamiltonian formulism of mechanics because here $t$
is independent of $x^0$. At this stage, therefore, we can not derive
the canonical momentum and Hamiltonian by means of $L,~x^i$ and
$\dot{x}^i$. If we choose the parameter $t$ such that
\begin{eqnarray}
ds = c dt
\end{eqnarray}
 and substitute eqs.(\ref{2.2}) into eqs.(\ref{2.3}), we
get the standard geodesic equation :
\begin{eqnarray}\label{2.4}
g_{\lambda \mu}{d^2 x^{\mu} \over ds^2}+\Gamma_{\lambda,\;
\mu\nu}{dx^\mu \over ds} {dx^\nu \over ds}=0,
\end{eqnarray}
where
\begin{eqnarray}
\Gamma_{\lambda, \; \mu\nu}=g_{\lambda\rho}\Gamma^\rho_{\mu\nu}
 = {1\over 2}(\pa_\mu g_{\lambda \nu}+\pa_\nu g_{\lambda\mu}-
 \pa_\lambda g_{\mu\nu}).
\end{eqnarray}

In order to derive the equations of motion in the Hamiltonian
framework, we have to fix the parameter
\begin{eqnarray}
    t=x^0/c~~~{\rm{or}}~~~x^0 =c t,
\end{eqnarray}
and write
\begin{eqnarray}
    S=\int L(x^i,\dot{x}^i,t) dt.
\end{eqnarray}
From eq.(\ref{2.2}), it is obvious that
\begin{eqnarray}\label{L}
    L(x^i,\dot{x}^i,t)=-m_0c^2 \sqrt{g_{00}(x^i,t)+2g_{0j}(x^i,t){1\over c}\dot{x}^j+g_{jk}(x^i,t) {1\over c^2}
    \dot{x}^j\dot{x}^k}
\end{eqnarray}
where $i$ only runs over the space indices. Then,  variating the
action with respect to both $x^i$ and t, we have the desired
Euler-Lagrangian equations  as follows
\begin{eqnarray}\label{star1}
  && {\pa L \over \pa x^i}={d \over dt} {\pa L \over \pa \dot{x}^i},
  \\ \label{star2}
  &&  {d \over dt}\left[ L-{\pa L \over \pa \dot{x}^i} \dot{x}^i\right]={\pa L \over \pa
    t}.
 \end{eqnarray}
These equations are the equations of motion in the time-dependent
Lagrangian-Hamiltonian framework, and the corresponding Lagrangian
can be used to deduce the momentum and energies of the system. It is
easy to check that under $x^0=ct$, Eqs.(\ref{star1}) (\ref{star2})
are consistent with 4-dimensional Euler-Lagrangian equation
(\ref{2.3})(or the geodesic equation (\ref{2.4})).

The equivalence of the two sets of equations comes from the fact
that the original action has a reparametrization symmetry of $t$ and
so the space and time coordinates are mixed together. That's to say,
\begin{eqnarray}
L={\pa L \over \pa \dot{x}^\lambda} \dot{x}^\lambda , \ \ \ \
 \lambda \ \mbox{runs over all the space and time indices}
\end{eqnarray}
which states that L is homogeneous of degree 1 as a function of
$\dot{x}^\lambda $. It's a special property of eq.(\ref{2.2}) but
also a general requirement for the action to have are
parametrization symmetry of t before $x^0$ is set to be $ct$. From
this we know that the above discussion for the equivalence of two
sets of Euler Lagrangian equations does not apply to the general
hamiltonian system but a special nice relation for the free particle
moving in the space and time described by theories of special
relativity.

\section{ Lagrangian, canonic momentum and Hamiltonian  of free particle in the de-Sitter invariant
special relativity }

According to the discussions in previous sections, similar to
 $L_{c}$ (see eq.(\ref{3})), the
Lagrangian for  free particle in $\mathcal{SR}_{cR}$ is
\begin{equation}\label{27}
 L_{{cR}}=-m_0c{ds\over
 dt}=-m_0c{\sqrt{B_{\mu\nu}(x)dx^\mu dx^\nu}\over dt}=-m_0c{\sqrt{B_{\mu\nu}(x)\dot{x}^\mu \dot{x}^\nu}},
 \end{equation}
where $\dot{x}^\mu=\frac{d}{dt}x^\mu$, $B_{\mu\nu}(x)$ is Beltrami
metric\cite{look,Lu74,Guo1,Guo2}:
\begin{eqnarray}\label{star28}
B_{\mu\nu}(x)={\eta_{\mu\nu} \over \sigma (x)}+{1\over R^2
\sigma(x)^2} \eta_{\mu\lambda}\eta_{\nu\rho} x^\lambda
x^\rho,~~~{\rm{with}}~~~\sigma(x)\equiv 1-{1\over R^2}
\eta_{\mu\nu}x^\mu x^\nu,
\end{eqnarray}
where the constant $R$ is the radius of the pseudo-sphere in {\it
dS}-space, and it can be related to cosmological constant via
$R=\sqrt{3/\Lambda}$\cite{Guo1}\cite{Guo2}. Seting up the time
$t=x^0/c$,   $B_{\mu\nu}(x)$ can be rewritten as follows
\begin{eqnarray}\label{star29}
ds^2&=&B_{\mu\nu}(x) dx^\mu dx^\nu
=\widetilde{g}_{00}d(ct)^2+\widetilde{g}_{ij}\left[(dx^i+N^id(ct))
(dx^j+N^jd(ct))\right]\\ \nonumber &=& c^2 (dt)^2
\left[\widetilde{g}_{00} +\widetilde{g}_{ij}({1\over
c}\dot{x}^i+N^i) ({1\over c}\dot{x}^j+  N^j)\right],
\end{eqnarray}
where
\begin{eqnarray}\label{star30}
\widetilde{g}_{00}&=&{R^2\over \sigma(x) (R^2-c^2t^2)},\\
\label{star31} \widetilde{g}_{ij}&=&{\eta_{ij}\over \sigma (x)}+
{1\over
R^2\sigma(x)^2}\eta_{il}\eta_{jm}x^lx^m,\\
\label{star32} N^i&=&{ctx^i \over R^2-c^2t^2}.
\end{eqnarray}
Substituting eqs.(\ref{star28})--(\ref{star32}) into (\ref{27}), we
obtain the Lagrangian for free particle in $ \mathcal{SR}_{cR} $:
\begin{equation}\label{33}
 L_{{cR}}=-m_0c^2 \sqrt{\widetilde{g}_{00} +\widetilde{g}_{ij}({1\over
c}\dot{x}^i+N^i) ({1\over c}\dot{x}^j+  N^j)}.
 \end{equation}
By means of the explicit expressions of
eqs.(\ref{star30})--(\ref{33}) and doing straightforward
calculations, we can prove the following equation:
\begin{equation}\label{initial motion condition}
 {\pa L_{{cR}}\over \pa x^i}= {\pa^2 L_{{cR}}\over \pa t \pa \dot{x}^i}
+ {\pa^2 L_{{cR}}\over \pa x^j \pa \dot{x}^i}\dot{x}^j.
 \end{equation}
Substituting eq.(\ref{33}) into the Euler-Lagrangian equation
(\ref{star1}) and using identity (\ref{initial motion condition}),
we have {\begin{equation}\label{35}
 {\pa^2 L_{{cR}}\over \pa \dot{x}^i \pa
 \dot{x}^j}\ddot{x}^j=-{m_0^4c^6R^4 \over
 L_{{cR}}^3 R^6 c^2 \sigma^{3}(x)}M_{ij}  \ddot{x}^j=0.
 \end{equation}}
where $M_{ij}$ is a matrix that satisfies $\det \left[ M_{ij}
\right]\neq 0$. We conclude
\begin{equation}\label{38}
\ddot{x}^j=0, ~~ \dot{x}^j={\rm constant}\
\end{equation}
{This result indicates that the free particle in the Beltrami
space-time $\mathcal{B} \equiv
\{x^\mu,\;g_{\mu\nu}(x)=B_{\mu\nu}(x)\}$ moves along straight line
and with constant coordinate velocities. Namely the inertial motion
law for free particles holds true in the space-time $\mathcal{B}$,
and hence the inertial reference frame can be set in $\mathcal{B}$.
Thus, by means of solving Euler-Lagrangian equations in the
Lagrangian-Hamiltonian formulation, we have reconfirmed the claim in
refs.\cite{look}\cite{Lu74} on the velocity of motion of
free-particles based on solving geodesic equation originally.

As an essential advantage in the Lagrangian-Hamiltonian formulation
over other formulism,  both canonical momentum $\pi_i$ conjugating
to the Beltrami-coordinate $x^i$ and  canonical energy $H_{cR}$ (or
Hamiltonian ) conjugating to the Beltrami-time $t$ for free
particles in the inertial reference frame can be determined
rationally by the mechanism principle. By the eq.(\ref{33}), the
canonical momentum and the canonical energy (or Hamiltonian) reads }
\begin{eqnarray}\label{40}
   \pi_{i} &=&\frac{\pa L_{cR}}{\pa
\dot{x}^i} = -m_0 \sigma(x) \Gamma B_{i \mu}\dot{x}^{\mu} \\
 \label{40a}    H_{cR}
&=&\sum_{i=1}^3 \frac{\pa L_{cR}}{\pa \dot{x}^i} \dot{x}^i
-L_{cR}=m_0 c \sigma(x) \Gamma B_{0 \mu}\dot{x}^{\mu}.
  \end{eqnarray}
where {\begin{eqnarray} \label{new parameter}
 \Gamma^{-1}\hskip-0.1in =\sigma(x) \frac{ds}{c dt}={1\over R} \sqrt{(R^2-\eta_{ij}x^i
x^j)(1+\frac{\eta_{ij}\dot{x}^i \dot{x}^j}{c^2})+2t \eta_{ij}x^i
\dot{x}^j -\eta_{ij}\dot{x}^i \dot{x}^j t^2+\frac{(\eta_{ij}
x^i\dot{x}^j)^2}{c^2}}.
\end{eqnarray}}
Under the equation of motion Eq.(\ref{38}), we have the following
relation
\begin{equation}\label{38a} \dot{\Gamma}|_{\ddot{x}^i=0}=0,
\end{equation}
whose corresponding one in $\mathcal{SR}_c$ is
\begin{equation}\label{38b}
\dot{\gamma}|_{\ddot{x}^i=0}\equiv{d\over dt}\left. \left({1\over
\sqrt{1-v^2/c^2}}\right)\right|_{v=constant}=0.
\end{equation}
It is easy to check that
\begin{equation}\label{38c}
\lim_{R\rightarrow \infty} \Gamma= \lim_{x^i\rightarrow 0} \Gamma
=\gamma\equiv \frac{1}{\sqrt{1-\frac{v^2}{c^2}}}.
\end{equation}
And,
% {Generally, canonical momenta $\pi_i$ and $H_{cR}$ are not the
%physical momentum and the energy of the particle respectively.
 in
the $R \to \infty$ limit, $\pi_i$ and $H_{cR}$ go back to the
standard Einstein Special Relativity's expressions:
\begin{equation} \label{44}
 \pi_{i}|_{R\rightarrow \infty}= \frac{m_0
 v_i}{\sqrt{1-\frac{v^2}{c^2}}},~~~~
    H_{cR}|_{R\rightarrow \infty} =\frac{m_0 c^2}
{\sqrt{1-\frac{v^2}{c^2}}}.
 \end{equation}
where $v_i=-\eta_{ij}\dot{x}^j$. Furthermore,  at the original point
of space-time coordinates $t=x^i=0$, but $R=$finite, we have also
expressions like (\ref{44}):
\begin{equation} \label{44-2}
 \pi_{i}|_{t=x^i=0}=\frac{m_0 v_i}{\sqrt{1-\frac{v^2}{c^2}}},~~~~ H_{cR}|_{t=x^i=0} =\frac{m_0 c^2}
{\sqrt{1-\frac{v^2}{c^2}}}.
 \end{equation}
In the Table I, we listed some results of Lagrange formulism  both
in the ordinary special relativity $\mathcal{SR}_c$ and in the de
Sitter invariant special relativity $\mathcal{SR}_{cR}$. Comparing
the results in $\mathcal{SR}_{cR}$ with ones in well known
$\mathcal{SR}_{c}$, we learned that as an extending theory of
$\mathcal{SR}_{c}$, $\mathcal{SR}_{cR}$ can simply be formulated by
a variable alternating in $\mathcal{SR}_{c}$: 1)
$\eta_{\mu\nu}\Rightarrow B_{\mu\nu}$; 2) $\gamma \Rightarrow \sigma
\Gamma$. This is a natural and nice feature for the Lagrangian
formulism of $\mathcal{SR}_{cR}$.

\begin{table}[hptb]
\begin{center}
\caption{Metric, Lagrangian, equation of motions, canonic momenta,
and Hamiltonian in the special relativity, $\mathcal{SR}_c$, and in
the de Sitter special relativity, $\mathcal{SR}_{cR}$. The
quantities $\gamma^{-1}=\sqrt{1+{\eta_{ij}\dot{x}^i \dot{x}^j \over
c^2}} $ and $\Gamma^{-1}\hskip-0.1in ={1\over R}
\sqrt{(R^2-\eta_{ij}x^i x^j)(1+\frac{\eta_{ij}\dot{x}^i
\dot{x}^j}{c^2})+2t \eta_{ij}x^i \dot{x}^j -\eta_{ij}\dot{x}^i
\dot{x}^j t^2+\frac{(\eta_{ij} x^i\dot{x}^j)^2}{c^2}} $ (see
eq.(\ref{new parameter})).}
\begin{tabular}{|c|c|c|}\hline\hline
    & $\mathcal{SR}_c$ & $\mathcal{SR}_{cR}$ \\ \hline
space-time metric & $\eta_{\mu\nu}$ & $B_{\mu\nu}(x),$
(Eq.(\ref{star28})) \\ \hline Lagrangian & $L_c=-m_0c^2\gamma^{-1}$
& $L_{cR}=-m_0c^2\sigma^{-1}\Gamma^{-1}$ \\ \hline equation of
motion & $v^i=\dot{x}^i=$constant, ( or $\dot{\gamma}=0$)
& $v^i=\dot{x}^i=$constant, ( or $\dot{\Gamma}=0$) \\
\hline canonic momenta & $\pi_i=-m_0\gamma \eta_{i\mu}\dot{x}^\mu $
& $\pi_i=-m_0\sigma \Gamma B_{i\mu}\dot{x}^\mu $ \\ \hline
Hamiltonian & $H_c=m_0c\gamma \eta_{0\mu} \dot{x}^\mu$ &
$H_{cR}=m_0c\sigma \Gamma B_{0\mu} \dot{x}^\mu$ \\ \hline \hline
\end{tabular}
\end{center}
\end{table}

\par
 Combining  Eq.(\ref{40}) with Eq.(\ref{40a}),
the covariant 4-momentum in $\cal{B}$ is:
\begin{equation}\label{46-1}
\pi_{\mu}\equiv (\pi_0,\;\pi_i)=(-\frac{H_{cR}}{c},\pi_i)=-m_0
\sigma \Gamma B_{\mu \nu} \dot{x}^{\nu}= -m_0 cB_{\mu
\nu}{d{x}^{\nu}\over ds},
\end{equation}
and
\begin{eqnarray}\label{dispersion}
B^{\mu\nu}\pi_{\mu}\pi_{\nu}&=&m_0^2 c^2.
\end{eqnarray}
From eqs.(\ref{33}) (\ref{40}) (\ref{40a}) (\ref{dispersion}), we
have
\begin{eqnarray}\label{hamilton}
H_{cR}=\sqrt{\widetilde{g}_{00}}\sqrt{m_0^2c^4
-c^2\widetilde{g}^{ij}\pi_i\pi_j}-c\pi_iN^i,
\end{eqnarray}
where $\widetilde{g}_{00},\;N^i$ have been shown in
eqs.(\ref{star30}) (\ref{star32}), and
$\widetilde{g}^{ij}={\sigma(x)(\eta^{ij}-\frac{x^i x^j}{R^2-c^2
t^2})}$ from eq.(\ref{star31}). It is straightforward to to get the
following canonical equations
\begin{eqnarray}\label{canonical equation}
\begin{array}{c}
     \dot{x}^i = \frac{\partial H_{cR}}{\partial \pi_i}=\{H_{cR}, x^i\}_{PB}  \\
     \dot{\pi}_i =-\frac{\partial H_{cR}}{\partial
x^i}=\{H_{cR}, \pi_i\}_{PB}.
  \end{array}
\end{eqnarray}
where the Poisson bracket
\begin{eqnarray}\label{Poisson}
\{x^i, \pi_j\}_{PB}=\delta^i_j, \;\;\{x^i,
x^j\}_{PB}=0,\;\;\{\pi_i, \pi_j\}_{PB}=0
\end{eqnarray}
 are as usual. It is also straightforward to check $\dot{x}^i=constant$ by
eq.(\ref{canonical equation}).

Finally, we would like to mention that generally, the canonical
momenta $\pi_i$ and the Hamiltonian $H_{cR}$ are not the physical
momentum and the energy of the particle respectively.

\section{Space-Time Symmetry of the de-Sitter invariant special relativity
and the Noether Charges }

{The space time transformations {preserving} the Beltrami metric
were discovered about {30 years ago} by Lu, Zhou and Guo
(LZG)\cite{look}\cite{Lu74}(see also Appendix).
   }When we transform from one initial Beltrami frame $ x^{\mu}$ to
another Beltrami frame $ \tilde{x}^{\mu}$, and {when } the origin
of the new frame is $a^{\mu}$ in the original frame, {the}
transformations between them with 10 parameters  is as follows
% law,which first derived by
%Lu-qikeng. The transformation are checked in the Appendix.

\begin{eqnarray}\label{transformation}
x^{\mu} \;-\hskip-0.10in\longrightarrow\hskip-0.4in^{LZG}
 ~~ \tilde{x}^{\mu} &=& \pm \sigma(a)^{1/2} \sigma(a,x)^{-1}
(x^{\nu}-a^{\nu})D_{\nu}^{\mu}, \\
    \nonumber D_{\nu}^{\mu} &=& L_{\nu}^{\mu}+R^{-2} \eta_{\nu
\rho}a^{\rho} a^{\lambda} (\sigma
(a) +\sigma^{1/2}(a))^{-1} L_{\lambda}^{\mu} ,\\
\nonumber L : &=& (L_{\nu}^{\mu})\in SO(1,3), \\
\nonumber \sigma(x)&=& 1-{1 \over R^2}{\eta_{\mu \nu}x^{\mu} x^{\nu}}, \\
\nonumber \sigma(a,x)&=& 1-{1 \over R^2}{\eta_{\mu \nu}a^{\mu}
x^{\nu}}.
\end{eqnarray}

{It will be called as LZG-transformation hereafter, and we prove
it in the Appendix by means of the method in \cite{Lu74}. Under
LZG-transformation, the $B_{\mu\nu}(x)$ and the action of
$\mathcal{SR}_{cR}$ transfer respectively as follows
\begin{equation} \label{B01}
 B_{\mu\nu}(x)\;-\hskip-0.10in\longrightarrow\hskip-0.4in^{LZG}
 ~~ ~\widetilde{B}_{\mu\nu}(\widetilde{x})={\pa x^\lambda \over \pa
 \widetilde{x}^\mu}{\pa x^\rho \over \pa
 \widetilde{x}^\nu}B_{\lambda\rho}(x)=B_{\mu\nu}(\widetilde{x}),
\end{equation}
\begin{equation} \label{B02}
 S_{cR}\equiv\int dt L_{cR}(t)=-m_0c\int dt {\sqrt{B_{\mu\nu}(x)dx^\mu dx^\nu} \over dt}
 \;-\hskip-0.05in-\hskip-0.05in\longrightarrow\hskip-0.4in^{LZG}
 ~ ~~\widetilde{S}_{cR}=S_{cR}.
\end{equation}
By the mechanics principle, this action {invariance indicates that}
there are 10 conserved Noether charges in $\mathcal{SR}_{cR}$ like
the $\mathcal{SR}_{c}$ case. For $\mathcal{SR}_{c}$ the Noether
charges are(e.g., see {\it pp581-586} and {\it Part 9} in
ref.\cite{Noether}):
\begin{eqnarray}\label{503a}
\begin{array}{rcl}
  &&{\rm{ Noether}\;charges\;for\;Lorentz\;boost:\;} ~~
 K_{c}^i=m_0 \gamma c (x^i- t \dot{x}^i) \\
 &&{\rm
Charges\;for\;space-transitions\;(momenta):}~~~  P_{c}^i=m_0 \gamma \dot{x}^i, \\
 &&{\rm Charge\;for\;time-transition\;(energy): }~~~
 E_{c}= m_0 c^2 \gamma \\
&&{\rm Charges\;for\;rotations\;in\;space\;(angular momenta):}~~~
L_{c}^i = \epsilon^{i}_{jk}x^{j}P^{k}.
\end{array}
\end{eqnarray}

Here $\gamma=\frac{1}{\sqrt{1-\frac{v^2}{c^2}}}$. Note the Noether
charges here are the same as the corresponding canonical quantities,
because the Lagrangian for $\mathcal{SR}_{c}$ is time-independent
and all the coordinates are cyclic. While in $\mathcal{SR}_{cR}$
there is no cyclic coordinates and the Lagrangian is space-time
dependent.

When  space rotations were neglected temporarily  for simplify, the
LZG-transformation both due to a Lorentz-like boost and a
space-transition in the $x^1$ direction with  parameters
$\beta=\dot{x}^1/c$ and $a^1$ respectively and due to a time
transition with  parameter $a^0$ can be explicitly written as
follows:
\begin{eqnarray}\label{general transformation}
\begin{array}{rcl}
t\rightarrow \tilde{t}&=& \frac{\sqrt{\sigma(a)}}{c \sigma(a,x)}
\gamma \left[ct-\beta x^1-a^0+ \beta a^1 +\frac{a^0-\beta
a^1}{R^2}\frac{a^0 ct-a^1 x^1-(a^0)^2 +(a^1)^2 }
{ \sigma(a)+\sqrt{\sigma(a)}} \right] \\
 x^1\rightarrow \tilde{x}^1&=& \frac{\sqrt{\sigma(a)}}{
\sigma(a,x)}\gamma \left[ x^1-\beta ct +\beta a^0 -a^1 +\frac{a^1-
\beta a^0}{R^2}
\frac{a^0 ct-a^1 x^1-(a^0)^2 +(a^1)^2}{ \sigma(a)+\sqrt{\sigma(a)}}\right]\\
 x^2\rightarrow
\tilde{x}^2&=&\frac{\sqrt{\sigma(a)}}{\sigma(a,x)}x^2 \\
 x^3\rightarrow
\tilde{x}^3&=&\frac{\sqrt{\sigma(a)}}{\sigma(a,x)}x^3
\end{array}
\end{eqnarray}
It is easy to check when $R\rightarrow \infty$ the above
transformation goes back to Poincar\'e transformation. Notice that
in the LZG-transformation there are 3 boost parameters
$\beta^{i}={\dot{x}^i \over c}=\frac{v^i}{c}$,  4 spacetime
transition parameters $(a^0,a^1,a^2,a^3)$ and 3 rotation parameters
$\theta^i=0$.  Here $(a^0,a^1,a^2,a^3)$ is the origin of the
resulting Beltrami initial frame
 in the original Beltrami frame.

In terms of the standard procedure (e.g.,see ref.\cite{Noether} {\it
pp.581-586} ), the Noether charges corresponding to  the LZG
transformation (eq.(\ref{transformation})) invariance can be derived
:
\begin{enumerate}
\item Space transitions:
\begin{eqnarray}\label{momenta}
\begin{array}{rcl}
 x^{\mu}\rightarrow \tilde{x}^{\mu} &=&x^{\mu}+\frac{\eta_{ij}
a^i
x^j}{R^2}x^{\mu}-a^i \delta^{\mu}_{i}\\
 G_{a}^{i} &=&-\frac{\pi_{\mu}x^{\mu}}{R^2}x^i{+}\eta^{ij}\pi_j =m_0
\Gamma \dot{x}^i.
\end{array}
\end{eqnarray}

\item Time transition charge:
\begin{eqnarray}\label{energy}
\begin{array}{rcl}
x^{\mu}\rightarrow \tilde{x}^{\mu} &=&x^{\mu}+\frac{ct
x^{\mu}}{R^2}a^0 -a^0 \delta^{\mu}_{0}\\  G_{a^{0}} &=& x^0
\frac{\pi_{\mu} x^{\mu}}{R^2}-\eta^{0\mu}\pi_{\mu}=m_0 c \Gamma.
\end{array}
\end{eqnarray}

\item Lorentz boost charges:
\begin{eqnarray} \label{boost charge}
\begin{array}{rcl}
x^{\mu}\rightarrow \widetilde{x}^{\mu}&=&(\gamma (ct-\beta x^1),\
\gamma(x^1-\beta c t),\  x^2,\  x^3)\\  G_{\beta}^{i} &=&- x^i \pi_0
-x^0 \pi_i = m_0 c \Gamma (x^i -t\dot{x}^i).
\end{array}
\end{eqnarray}

\item Rotation charges:
\begin{eqnarray} \label{rotaztion charge}
\begin{array}{rcl}
x^{\mu}\rightarrow \tilde{x}^{\mu} &=&(ct, x^1+\theta x^2,
x^2-\theta x^1, x^3)\\  G_{\omega}^{i} &=&-m_0 \Gamma \ep^{i}_{\;jk}
x^j \dot{x}^k.
\end{array}
\end{eqnarray}
\end{enumerate}

Some remarks are in order:
\begin{enumerate}

\item{ Since a free particle which moves with a constant speed
$\ddot{x}^i=0$, we have already proved (see (\ref{38a}))
\begin{equation}\label{condition}
\frac{d}{dt}\Gamma=0.
\end{equation}
By using (\ref{condition}), we can check that those charges derived
above are indeed conservative:
\begin{eqnarray}\label{hehe}
\frac{d}{dt}G_{\lambda}|_{\lambda=a^0,a^i,\beta^i,\omega^i} &=&0.
\end{eqnarray}
}

\item {In the limit that $R\rightarrow\infty$ the Noether charges in $\mathcal{SR}_{cR}$
are the same as those in $\mathcal{SR}_{c}$, see eq.(\ref{503a}) }

\item {The mechanical (or physical) momenta and energy in the
Lagrangian-Hamiltonian formalism are defined as the Noether charges
corresponding to the space transitions, therefore the particle's
momenta and energy in $\mathcal{SR}_{cR}$ read
\begin{eqnarray}\label{physical momenta}
p_{cR}^i & \equiv &G_{a}^{i} = m_0 \Gamma \dot{x}^i
\\ \label{physical momenta 1}
E_{cR}& \equiv & {cp_{cR}^0}=c G_{a^{0}} =  m_0 c^2 \Gamma.
\end{eqnarray}

which are conservative quantities. We address that distinguishing
from the $\mathcal{SR}_{c}$, in $\mathcal{SR}_{cR}$ the physical
momentum $p^i$ of the particle is different from its canonical
momentum $\pi_i$. The former is conservative and the latter is
space-time-dependent. Combining eq.(\ref{physical momenta}) with
(\ref{physical momenta 1}), we have the 4-momentum in
$\mathcal{SR}_{cR}$ as follows
\begin{eqnarray}\label{70-p4}
p_{cR}^\mu \equiv \{p_{cR}^0,~p_{cR}^i\} = m_0 \Gamma \dot{x}^\mu
={m_0c\over \sigma(x)}{dx^\mu \over ds}=-{1\over
\sigma(x)}B^{\mu\nu}\pi_\nu,
\end{eqnarray}
which is consistent with energy-momentum definition in
ref.\cite{Lu74}.}

\item{
In general the boost Noether charges for $\mathcal{SR}_{cR}$ are
\begin{eqnarray} \label{63-g}
K_{cR}^{i} \equiv G_{\beta}^{i} = x^{i}p_{cR}^{0}+x^{0}p_{cR}^{i} =
m_0 c \Gamma (x^i -t\dot{x}^i) .
\end{eqnarray}
While the the angular momentum  are
\begin{eqnarray} \label{75-G}
L_{cR}^{i} \equiv G_{\omega}^{i} =-m_0 \Gamma \ep^{i}_{\;jk} x^j
\dot{x}^k.
\end{eqnarray}

From eq.(\ref{70-p4}) we have the dispersion relation
\begin{eqnarray}\label{diff}
\sigma^2(x)B_{\mu \nu}p_{cR}^\mu p_{cR}^\nu = m_{0}^2 c^2,
\end{eqnarray}
And we can rewrite it using the Noether charges
\begin{eqnarray} \label{sesan}
E_{cR}^2 =m_0^2 c^2+{\mathbf p}_{cR}^2 + \frac{c^2}{R^2} ({\mathbf
L}_{cR}^2-{\mathbf K}_{cR}^2).
\end{eqnarray}
Here $E_{cR},{\mathbf p}_{cR},{\mathbf L}_{cR},{\mathbf K}_{cR}$ are
conserved physical energy, momentum, angular-momentum and boost
charges respectively. }
\end{enumerate}

\section{quantum mechanics for one particle in $\mathcal{SR}_{cR}$}

Lagrangian-Hamiltonian formulation of mechanics is the foundation of
quantization. When the classical Poisson brackets in canonical
equations for canonical coordinates and canonical momentum become
operator's commutators, i.e., $\{x,\pi\}_{PB}\Rightarrow {1\over
i\hbar}[x,\hat{\pi}]$, the classical mechanics will be quantized. In
this way, for instance, the ordinary relativistic (i.e.,
$\mathcal{SR}_c$) one-particle quantum equations have been derived.
To the particle with spin-0, that is just the well known
Klein-Gordon equation.  In the canonic quantization formulism for
$\mathcal{SR}_{cR}$, the canonic variable operators are $x^i, \;
\hat{\pi}_i $ with $i=1,2,3  $, and due to eq.(\ref{Poisson}) the
basic commutators for the free particle quantization theory of
$\mathcal{SR}_{cR}$ are the same as usual, i.e.,
\begin{eqnarray}\label{ij}
[x^i, \;\hat{\pi}_j]=i\hbar \delta^i_j,~~~~~ [\hat{\pi}_i,
\;\hat{\pi}_j]=0, ~~~~~[x_i, \;x_j]=0.
\end{eqnarray}
The Hamiltonian operator $\hat{H}_{cR}\equiv -c\hat{\pi}_0$
represents the generator of time evolution, i.e.,
\begin{equation}\label{time}
[t,\;\hat{H}_{cR}]=-i\hbar~,~~~~{\rm
or}~~~[x^0,\;\hat{\pi}_0]=i\hbar~.
\end{equation}
 Since the time evolution is independent of the space coordinate
displacements whose generators are $\hat{\pi}_i$, we always have
\begin{equation}\label{pi-0}
[\hat{H}_{cR},\;\hat{\pi}_i]=0~,~~~~{\rm
or}~~~[\hat{\pi}_{0},\;\hat{\pi}_i]=0~,
\end{equation}
which is independent of the dynamics ( or the dispersion
relation)\cite{foot}. Combining eqs. (\ref{ij}), (\ref{time}) and
(\ref{pi-0}), we have (hereafter the hat notations for operators are
removed):

%The Hamiltonian operator $\hat{H}_{cR}\equiv
%-c\hat{\pi}_0=f(t,x^i,\hat{\pi}^i)$ with the $c$-number limit
%$f(t,x^i,\hat{\pi}^i)|_{\hat{\pi}\Rightarrow \pi}= H_{cR}$. It is
%well know that due to terms $\widetilde{g}_{00}^2(t,x^i)
%\widetilde{g}^{ij}(t,x^i)\pi_i\pi_j$ and $\pi_iN^i(t,x^i)$ emerging
% in $H_{cR}$ (see eq.(\ref{hamilton})), the ordering of $x^i$ and
%$\hat{\pi}^i$ has to be taken care. Generally, the most symmetrical
%ordering (i.e., Weyl ordering) is favored for realistic quantization
%scheme.

  }

\begin{eqnarray}\label{quantum}
[x^{\mu},\pi_{\nu}]&=&i\hbar\delta^{\mu}_{\nu}, \\ \nonumber
[x^{\mu},x^{\nu}]&=& 0, \\ \nonumber [\pi_{\mu},\pi_{\nu}]&=& 0.
\end{eqnarray}
{ The general solution of eq.(\ref{quantum}) is
\begin{equation}\label{solution11}
\pi_\mu = -i\hbar \pa_\mu +(\pa_\mu G(t,x)),
\end{equation}
where $G(t,x)$ is a function of $t$ and $x^i$. Now, the dynamical
Hamiltonian  ${H}_{cR}\equiv -c \pi_0$ is $(\pi x)$-product term
dependent (see eq.(\ref{hamilton})), and  the ordering of $x^i$ and
$\pi^i$ has to be taken care. Generally, the most symmetrical
ordering (i.e., Weyl ordering) is favored for realistic quantization
scheme. To $\mathcal{SR}_{cR}$, we prefer the quantization scheme
that protects the de Sitter symmetry $SO(1,4)$. This requirement
will lead to fix the function $G(t,x)$ in eq.(\ref{solution11}). By
this consideration,
 we take\cite{path} }
\begin{eqnarray}\label{quantum operator}
\pi_{\mu} &=&-i
\hbar\acute{D}_{\mu}~=-i\hbar(\pa_{\mu}+\frac{\Gamma_{\mu}}{2})
=-i\hbar B^{-\frac{1}{4}}\pa_{\mu} B^{1\over 4},
\end{eqnarray}
where $B=det(B_{\mu \nu}),\;\Gamma_{\mu}=\Gamma^{\nu}_{\mu\nu}$. {
Eq.(\ref{quantum operator}) indicates $G(t,x)=-i\hbar \;\log
(B^{1\over 4}).$ In contrast with the ordinary quantization
discussions to the theories in curved space only\cite{path}, our
treatment presented here is suitable for the theories in generic
curved space-time, in which the 4-dimensional metric is time- and
space-dependent. The classical dispersion relation
(\ref{dispersion}) can be rewritten as symmetric version $
B^{-\frac{1}{4}}\pi_{\mu}B^{\frac{1}{4}} B^{\mu\nu}
B^{\frac{1}{4}}\pi_{\nu}B^{-\frac{1}{4}}=m_{0}^2 c^2  $, and then
the $\mathcal{SR}_{cR}$-one particle wave equation reads
\begin{eqnarray}\label{KG}
B^{-\frac{1}{4}}\pi_{\mu}B^{\frac{1}{4}} B^{\mu\nu}
B^{\frac{1}{4}}\pi_{\nu}B^{-\frac{1}{4}}\phi(x,t)&=&m_{0}^2 c^2
\phi(x,t)\;,
\end{eqnarray}
where $\phi(x,t)$ is the particle's wave function.  Substituting
(\ref{quantum operator}) into (\ref{KG}), we have
\begin{equation}\label{KG in curved spactime}
\frac{1}{\sqrt{B}}\pa_{\mu}(B^{\mu\nu}\sqrt{B}\pa_{\nu})\phi+\frac{m_{0}^2
c^2}{\hbar^2} \phi=0,
\end{equation}
which is just the Klein-Gordon equation in  curved space-time with
Beltrami metric $B_{\mu\nu}$, and its explicit form is
\begin{equation}\label{KG1}
(\eta^{\mu\nu}-\frac{x^{\mu}x^{\nu}}{R^2})\partial_{\mu}\partial_{\nu}\phi
- 2\frac{x^{\mu}}{R^2}\partial_{\mu}\phi+\frac{m_{0}^2c^2}{\hbar^2
\sigma(x)} \phi = 0,
\end{equation}
which is the desired $\mathcal{SR}_{cR}$-quantum mechanics equation
for free particle.

Substituting (\ref{quantum operator}) into }(\ref{70-p4}), we obtain
the physical momentum and energy operators (noting the subscripts
$cR$ for $p_{cR}^\mu,\;L_{cR}^{\mu\nu}$ in (\ref{70-p4}) will be
moved hereafter):
\begin{eqnarray}\label{momentum operator}
p^{\mu} &=& i\hbar
[(\eta^{\mu\nu}-\frac{x^{\mu}x^{\nu}}{R^2})\partial_{\nu}+\frac{5x^{\mu}}{2R^2}].
\end{eqnarray}
$p^\mu$ together with  operator $ L^{\mu\nu} = x^{\mu} p^{\nu}
-x^{\nu} p^{\mu}$ form a  algebra as follows
\begin{eqnarray} \label{ds algebra}
[p^{\mu},p^{\nu}]&=&\frac{1}{R^2}L^{\mu\nu}
\\ \nonumber
[L^{\mu\nu},p^{\rho}]&=&\eta^{\nu\rho}p^{\mu}-\eta^{\mu\rho}p^{\nu}\\
\nonumber [L^{\mu\nu},L^{\rho\sigma}]&=&\eta^{\nu\rho}L^{\mu\sigma}-
\eta^{\nu\sigma}L^{\mu\rho}+\eta^{\mu\sigma}L^{\nu\rho}-\eta^{\mu\rho}L^{\nu\sigma}
\end{eqnarray}
which is just the de-Sitter algebra SO(1,4). { This fact means that
the quantization scheme presented in this paper preserves the
external space-time symmetry of $\mathcal{SR}_{cR}$. }

\section{Summery and Discussions}
{In this paper, we have provided a systemic study to the de Sitter
invariant special relativity with Beltrami metric in terms of the
Lagrangian-Hamiltonian formulism.} In this theory there are two
universal parameters $c$ and $R$, and it was denoted as
$\mathcal{SR}_{cR}$. Distinguishing from the Minkowski metric
$\eta_{\mu\nu}$, the Beltrami metric is space-time dependent.
Therefore the principle of least action for space-time dependent
Lagrangian is reexamined in order to be sure the Lagrangian equation
is consistent with the geodesic equation in Beltrami space-time
$\mathcal{B}$. Following standard procedure in the ordinary special
relativity and by means of the Beltrami metric we construct the
Lagrangian $L_{cR}(t, \mathbf{x}, \mathbf{\dot{x}})$ for
$\mathcal{SR}_{cR}$. The inertial law has been reconfirmed in
$\mathcal{SR}_{cR}$ by means of solving its equation of motion in
the Lagrangian-Hamiltonian formulism, which leads to well {defined}
inertial coordinate reference frame in Beltrami space-time
$\mathcal{B}$. The canonic momenta and canonic energy (or
Hamiltonian) are derived. It is found that both of them are
space-time dependent, which is due to that there are no cyclic
coordinates in $L_{cR}$ and the $L_{cR}$ is time dependent. The
canonic equations and {the corresponding Poisson bracket expressions
are obtained.} The canonic formulation is useful for quantization of
the mechanics in $\mathcal{SR}_{cR}$. The de Sitter transformation
in space-time $\mathcal{B}$ (i.e., LZG-transformation) has been used
to derive the Noether charges of $\mathcal{SR}_{cR}$. Ten
conservative charges are obtained. They are: 3 boost charges, 4
momentum-energy charges and 3 angular momentum charges. In this way
and by the symmetry principle, the physical momenta, the physical
energy and the physical angular momenta in $\mathcal{SR}_{cR}$ are
determined in the Lagrangian-Hamiltonian formulism. It has been
found that the Hamiltonian is not equal to the energy, and the
canonical momentum is also different from the physical momentum,
i.e., $H\neq E$ and $\overrightarrow{\pi}\neq \overrightarrow{p}$.
This is a significant property for $\mathcal{SR}_{cR}$. When
$R\rightarrow \infty$, all results of the de Sitter invariant
special relativity goes back to the ordinary special relativity.

{ By means of the canonic formulation, the quantum mechanics of
$\mathcal{SR}_{cR}$ is achieved. The one particle quantum equation
is just the Klein-Gordon equation in  curved space-time with
Beltrami metric $B_{\mu\nu}$. The quantization scheme with proper
$(\pi-x)$-ordering preserves the external space-time symmetry of
$\mathcal{SR}_{cR}$. When $R\rightarrow \infty$ or  $x \rightarrow
0$, the theory goes back to} { ordinary one particle quantum
equation of the Einstein's special relativity, i.e., the ordinary
Klein-Gordon equation in flat space-time. A further discussion on
the solutions of the equation of $\mathcal{SR}_{cR}$-quantum
mechanics would be interesting, which, however, is left to be in our
coming works. }

Physically, since $R$ in the $\mathcal{SR}_{cR}$ could be a very
large distance parameter, say the "radius of universe horizon", the
existing experiments can not justify or  rule out
$\mathcal{SR}_{cR}$. Therefore,  how to design experiments to detect
the effects of the de Sitter invariant special relativity would be
remarkable. { We speculate that careful studies on the solutions of
$\mathcal{SR}_{cR}$-quantum mechanics may bring us ideas for this
aim. For instance, the master equation for the photons emitted from
very far away star should be the equation of
$\mathcal{SR}_{cR}$-quantum mechanics eq.(\ref{KG1}) with $m_0=0$
instead of the ordinary KG-equation of $\mathcal{SR}_{c}$, because
the distance $|x|\sim R$. This difference may lead to reveal some
effects to distinguish $\mathcal{SR}_{cR}$ from $\mathcal{SR}_{c}$.
}

 { Finally, we would like to briefly mention the
 Double Special Relativity (DSR)\cite{DSR} in comparison with
 $\mathcal{SR}_{cR}$. DSR is an interesting theory, and  is another
modified special relativity with also two universal constants: $c$
and Planck length $l_P\equiv \sqrt{\hbar G_N/c^3} $ (or a length
$l=\hbar /(\kappa c)$ near $l_P$, where $\kappa \sim m_P\equiv
\sqrt{\hbar c/G_N}$ ). Obviously, the length parameter of DSR is
drastically smaller than one of $\mathcal{SR}_{cR}$: $l_P/R \sim
10^{-120}$. This indicates that the physics discussed in DSR is very
different from one in $\mathcal{SR}_{cR}$: DSR is inspired by
quantum gravity and by a space-time quantization treatment for over
the ultraviolet tragedy in quantum field theory\cite{snyder}, while
$\mathcal{SR}_{cR}$ is motived by naturally extending the space-time
and the dynamics theory of Einstein's special relativity
$\mathcal{SR}_{c}$, and the corresponding remarkable physics is
related to the cosmology, say the propagation of photons emitted
from far away stars with distance $|x|\sim R$. In other words,  like
$\mathcal{SR}_{c}$, $\mathcal{SR}_{cR}$  preserves a specific
space-time metric} { (i.e., $B_{\mu\nu}$) and the inertial frames
are well defined. And then like $\mathcal{SR}_{c}$ further,
$\mathcal{SR}_{cR}$  has  well defined Lagrangian-Hamiltonian
formulation too. Consequently, a consistent quantum mechanics of
$\mathcal{SR}_{cR}$ exists and can be derived by means of the
standard quantization procedures relied on the first principle of
quantum theory. However, the models of DSR are all different from
$\mathcal{SR}_{cR}$ in these aspects. Basically,  DSR theories can
be understood as particular realizations of deformed $\kappa-$
Poincar\'{e} algebra in momentum spaces\cite{DSR2}, or of a de
Sitter geometry in momentum space\cite{DSR3}. Due to this structure,
the space coordinates in DSR are non-commutative, i.e.,
$[\hat{x}^\mu,\;\hat{x}^\nu]\neq 0 $, (which is in conflict with the
principle requirement of quantum mechanics (see eqs.(\ref{ij})
(\ref{quantum}))), and hence there are no any Lagrangian-Hamiltonian
formulations for DSR yet, which can be constructed consistently. If
the length scale for both DSR and $\mathcal{SR}_{cR}$ were denoted
as $\mathcal{R}$, then DSR is a theory for $|x|> \mathcal{R}(\equiv
l_P)$, and $\mathcal{SR}_{cR}$ is for $|x|< \mathcal{R}(\equiv R)$
(see eq.(\ref{A11}). Therefore, there is no overlapping part for DSR
and $\mathcal{SR}_{cR}$, and the theory structures of two theories
must be independent each other. }

\begin{center} {\bf ACKNOWLEDGMENTS}
\end{center}
One of us (MLY) would like to acknowledge Professor Han-Ying Guo,
Qi-Keng Lu, and Yong-Shi Wu for stimulation discussions. The
authors  also thank  Dr Shao-Xia Chen. This work is partially
supported by National Natural Science Foundation of China under
Grant Numbers 90403021, and by the PhD Program Funds of the
Education Ministry of China under Grant Number 20020358040.

%\newpage
\begin{appendix}
\section{Space-Time Transformation to Preserve Beltrami Metric}
{\small
%\noindent

{ Now we prove that under the LZG space-time transformation
eq.(\ref{transformation}) in the text the Beltrami metric is
invariant.}

We define the field $\mathfrak{D}_{\lambda}(m,n)$ to be all $m\times
n$ matrix X such that
\begin{eqnarray}\label{D}
   I-\lambda XJX^{\prime} >0
\end{eqnarray}
Here, I is the $m\times m$ identity matrix, J= diag[1,-1,...,-1] is
$n\times n$ matrix, $\lambda=1/R^2\neq 0$ is a real number. A real
matrix $A>0$ means that A is positive definite. Let A, B, C, D be
$m\times m$, $n\times m$, $m\times n$, $n\times n$ matrices
respectively, satisfying
\begin{eqnarray}\label{trans}
    \left( \begin{matrix}A & C\\ B & D \end{matrix}
    \right)\left(\begin{matrix}I & 0\\0& -\lambda J \end{matrix}\right)\left( \begin{matrix}A & C\\B & D
    \end{matrix}\right)^{\prime}=\left(\begin{matrix}I & 0\\0& -\lambda J\end{matrix}\right)
\end{eqnarray}
write out the entries we get
\begin{eqnarray}\label{trans1}
    AA^{\prime}-\lambda CJC^{\prime}=I,\ AB^{\prime}=\lambda
    CJD^{\prime},\ BB^{\prime}-\lambda DJD^{\prime}=-\lambda J
\end{eqnarray}
Eqs.(\ref{trans}) is also equivalent to
\begin{eqnarray*}
    &&\left( \begin{matrix}A & C\\ B & D \end{matrix}
    \right)\left(\begin{matrix}I & 0\\0& -\lambda J \end{matrix}\right)\left( \begin{matrix}A & C\\B & D
    \end{matrix}\right)^{\prime}\left(\begin{matrix}I & 0\\0& -\lambda^{-1}
    J\end{matrix}\right)=\left( \begin{matrix}I & 0\\ 0 & I
    \end{matrix}\right)\\
  \Leftrightarrow &&\left(\begin{matrix}I & 0\\0& -\lambda J \end{matrix}\right)\left( \begin{matrix}A & C\\B & D
    \end{matrix}\right)^{\prime}\left(\begin{matrix}I & 0\\0& -\lambda^{-1}
    J\end{matrix}\right)\left( \begin{matrix}A & C\\ B & D \end{matrix}
    \right)=\left( \begin{matrix}I & 0\\ 0 & I
    \end{matrix}\right)\\
    \Leftrightarrow&&\left( \begin{matrix}A & C\\B & D
    \end{matrix}\right)^{\prime}\left(\begin{matrix}I & 0\\0& -\lambda^{-1}
    J\end{matrix}\right)\left( \begin{matrix}A & C\\ B & D \end{matrix}
    \right)=\left( \begin{matrix}I & 0\\ 0 & -\lambda^{-1}J
    \end{matrix}\right)\\
    \Leftrightarrow &&\left( \begin{matrix}A & C\\B & D
    \end{matrix}\right)^{\prime}\left(\begin{matrix}\lambda I & 0\\0& -
    J\end{matrix}\right)\left( \begin{matrix}A & C\\ B & D \end{matrix}
    \right)=\left( \begin{matrix}\lambda I & 0\\ 0 & -J
    \end{matrix}\right)
\end{eqnarray*}
write out the entries we get
\begin{eqnarray}\label{trans2}
    \lambda A^{\prime}A-B^{\prime}JB=\lambda I,\ \lambda
    A^{\prime}C=B^{\prime}JD,\ \lambda C^{\prime}C- D^{\prime}JD=-J
\end{eqnarray}
Therefore, (\ref{trans1}) and (\ref{trans2}) are equivalent. Observe
that (\ref{D}) can be written as
\begin{eqnarray}
    \left(\begin{matrix}I & X\end{matrix}\right) \left( \begin{matrix} I & 0\\ 0 & -\lambda J
    \end{matrix}\right)
    \left(\begin{matrix}I\\X^{\prime}\end{matrix}\right)>0
\end{eqnarray}
we use (\ref{trans}) and get
\begin{eqnarray*}
&&\left(\begin{matrix}I & X\end{matrix}\right) \left(
\begin{matrix}A & C\\ B & D \end{matrix}
    \right)\left(\begin{matrix}I & 0\\0& -\lambda J \end{matrix}\right)\left( \begin{matrix}A & C\\B & D
    \end{matrix}\right)^{\prime}
    \left(\begin{matrix}I\\X^{\prime}\end{matrix}\right)>0\\
   \Leftrightarrow &&(A+XB) \left(\begin{matrix}I & Y
   \end{matrix}\right)
   \left(\begin{matrix}I & 0\\0& -\lambda J \end{matrix}\right)
    \left(\begin{matrix}I\\Y^{\prime}\end{matrix}\right)(A+XB)^{\prime}>0\\
    \Leftrightarrow &&\left(\begin{matrix}I & Y
   \end{matrix}\right)
   \left(\begin{matrix}I & 0\\0& -\lambda J \end{matrix}\right)
    \left(\begin{matrix}I\\Y^{\prime}\end{matrix}\right)>0
\end{eqnarray*}
here
\begin{eqnarray}\label{Y}
    Y=(A+XB)^{-1}(C+XD)
\end{eqnarray}
Therefore the transformation (\ref{Y}) maps
$\mathfrak{D}_{\lambda}(m,n)$ to itself and is an automorphism. We
can define a metric on $\mathfrak{D}_{\lambda}(m,n)$
\begin{eqnarray}\label{metric}
    ds^2=tr \left\{ (I-\lambda XJX^{\prime})^{-1} dX (J-\lambda X^{\prime}X)^{-1}dX^{\prime} \right\}
\end{eqnarray}

\noindent \textbf{We claim that this metric is invariant under the
transformation}
(\ref{Y})\\
\noindent \textbf{Proof}: Note that from the above discussion we
have
\begin{eqnarray}
    (I-\lambda XJX^{\prime})=(A+XB)(I-\lambda YJY^{\prime})(A+XB)^{\prime}
\end{eqnarray}
since $X=(AY-C)(D-BY)^{-1}$, we also have
\begin{eqnarray*}
    && J-\lambda Y^{\prime}Y\\
    &=& \left(\begin{matrix}Y^{\prime} & -I\end{matrix}\right) \left( \begin{matrix} -\lambda I & 0\\ 0 & J
    \end{matrix}\right)
    \left(\begin{matrix}Y\\-I\end{matrix}\right)\\
    &=& \left(\begin{matrix}Y^{\prime} & -I\end{matrix}\right) \left( \begin{matrix}A & C\\B & D
    \end{matrix}\right)^{\prime}\left(\begin{matrix}-\lambda I & 0\\0&
    J\end{matrix}\right)\left( \begin{matrix}A & C\\ B & D \end{matrix}
    \right)
    \left(\begin{matrix}Y\\-I\end{matrix}\right)\\
    &=&(D-BY)^{\prime}  \left(\begin{matrix}X^{\prime} & -I\end{matrix}\right) \left(\begin{matrix}-\lambda I & 0\\0&
    J\end{matrix}\right)
    \left(\begin{matrix}X\\-I\end{matrix}\right) (D-BY)\\
    &=&(D-BY)^{\prime} (J-\lambda X^{\prime}X)(D-BY)
\end{eqnarray*}
\begin{eqnarray*}
    dY&=&d((A+XB)^{-1}(C+XD))\\
      &=&\left[ -(A+XB)^{-1}d(A+XB)(A+XB)^{-1}(C+
      XD)+(A+XB)^{-1}d(C+XD) \right]\\
      &=&\left[ -(A+XB)^{-1}dXBY+(A+XB)^{-1}dXD \right]\\
      &=&(A+XB)^{-1}dX(D-BY)\\
      \\
    dY^{\prime}&=&d((A+XB)^{-1}(C+XD))^{\prime}=(D-BY)^{\prime}dX^{\prime}(A+XB)^{\prime -1}
\end{eqnarray*}
hence
\begin{eqnarray*}
    &&tr \left\{ (I-\lambda YJY^{\prime})^{-1} dY (J-\lambda Y^{\prime}Y)^{-1}dY^{\prime}
    \right\}\\
        &=&tr \{(A+XB)^{\prime}(I-\lambda
        XJX^{\prime})^{-1}(A+XB)(A+XB)^{-1}dX(D-BY)\\
       &&(D-BY)^{-1}(J-\lambda X^{\prime}X)^{-1}(D-BY)^{\prime -1}
       (D-BY)^{\prime}dX^{\prime}(A+XB)^{\prime -1}\}\\
       &=&tr \left\{ (I-\lambda XJX^{\prime})^{-1} dX (J-\lambda X^{\prime}X)^{-1}dX^{\prime} \right\}
\end{eqnarray*}
which  states that the metric (\ref{metric}) is invariant under
transformation (\ref{Y}). \hfill $\square$
\\
If we let $X_0=-CD^{-1}$, then
\begin{eqnarray*}
    Y&=&(A+XB)^{-1}(C+XD)=A^{-1}(I+XBA^{-1})^{-1}(X+CD^{-1})D
\end{eqnarray*}
The conditions in (\ref{trans2}) are equivalent to the following
\begin{eqnarray*}
    &&BA^{-1}=(\lambda CD^{-1}J)^{\prime}=\lambda J X_0^{\prime}\\
    &&(AA^{\prime})^{-1}=A^{\prime
    -1}(A^{\prime}A-\lambda^{-1}B^{\prime}JB)A^{-1}=(I-\lambda X_0
    JX_0^{\prime})\\
    &&(DJD^{\prime})^{-1}=D^{\prime -1}J D^{-1}=D^{\prime -1}(D^{\prime}JD-\lambda C^{\prime}C)D^{-1}
    =J-\lambda X_0^{\prime}X_0
\end{eqnarray*}
We get the formula
\begin{eqnarray}\label{2a}
    Y=A^{-1}(I-\lambda XJX_0)^{-1}(X-X_0)D
\end{eqnarray}
where the matrices A, D satisfy
\begin{eqnarray}
    AA^{\prime}=(I-\lambda X_0 J X_0^{\prime})^{-1},\
    DJD^{\prime}=(J-\lambda X_0^{\prime}X_0)^{-1}
\end{eqnarray}
\\
\\
For the special case $\mathfrak{D}_{\lambda}(1,4)$ in our paper,
$X=(X^0,X^1,X^2,X^3)$, $\mathfrak{D}_{\lambda}(1,4)$ is just
\begin{eqnarray}\label{A11}
    1-\lambda\eta_{\mu\nu}x^\mu x^\nu >0
\end{eqnarray}
The metric (\ref{metric}) now takes the form
\begin{eqnarray}
    ds^2&=&{dX(J-\lambda X^{\prime}X)^{-1}dX^{\prime}\over 1-\lambda
    XJX^{\prime}} =g_{\mu\nu}dx^\mu dx^\nu \\ \nonumber
    g_{\mu\nu}&=&{\eta_{\mu\nu}\over 1-\lambda \eta_{\lambda\rho}x^\lambda x^\rho}
    +{\lambda \eta_{\mu\lambda}\eta_{\nu\rho}x^\lambda x^\rho \over (1-\lambda
    \eta_{\alpha\beta}x^\alpha x^\beta)^2}
\end{eqnarray}
{Comparing eq.(\ref{metric}) with eq.(\ref{star28}) in the text, we
see $g_{\mu\nu}$ is just the Beltrami metric, i.e.,
$g_{\mu\nu}=B_{\mu\nu}(x)$. } By our claim, this metric is invariant
under the transformation (\ref{2a}), which now becomes
\begin{eqnarray}\label{A14}
    y^\mu=\sqrt{1-\lambda\eta_{\lambda\rho}a^\lambda a^\rho}{(x^\nu-a^\nu)D^\mu_\nu\over
    1-\lambda \eta_{\alpha\beta}a^\alpha x^\beta}
\end{eqnarray}
where we denote $X_0=(a^0,a^1,a^2,a^3)$ and $\{D^\mu_\nu\}$ are
constants, satisfying
\begin{eqnarray}\label{A15}
    \eta_{\lambda\rho}D^\lambda_\mu D^\rho_\nu=\eta_{\mu\nu}+{\lambda \eta_{\mu\lambda}\eta_{\nu\rho}a^\lambda a^\rho
    \over 1-\lambda \eta_{\alpha\beta}a^\alpha a^\beta}.
\end{eqnarray}
{By using the notations in the text:
$\widetilde{x}^\mu=y^\mu,\;\;\sigma(x)=1-\lambda
\eta_{\alpha\beta}x^\alpha x^\beta,\;\;\sigma(a,x)=1-\lambda
\eta_{\alpha\beta}a^\alpha x^\beta $, we rewrite eqs. (\ref{A14})
(\ref{A15}) as follows
\begin{eqnarray}\label{A16}
    \widetilde{x}^\mu&=&\sqrt{\sigma(a)}{(x^\nu-a^\nu)D^\mu_\nu\over
    \sigma(a,x)}~~~~,\\
\label{A17} \eta_{\lambda\rho}D^\lambda_\mu D^\rho_\nu &=&
\eta_{\mu\nu}+{\lambda \eta_{\mu\lambda}\eta_{\nu\rho}a^\lambda
a^\rho \over \sigma(a)}.
\end{eqnarray}
Taking ansatz
\begin{eqnarray}\label{A18}
    D^\mu_\nu=\pm (L^\mu_\nu+A \lambda \eta_{\nu\lambda}a^\lambda a^\rho
L^\mu_\rho)
\end{eqnarray}
 where $A$ is a constant which is determined by the normalization
constraint eq.(\ref{A17}):
\begin{eqnarray}\label{A19}
    A={1\over \sigma(a)+\sqrt{\sigma}}.
\end{eqnarray}
Substituting eqs.(\ref{A17})(\ref{A18})(\ref{A19}) into
eq.(\ref{A16}), we finally obtain
\begin{eqnarray}\label{A20}
    \widetilde{x}^\mu=\pm
    \sqrt{\sigma(a)}\sigma(a,x)^{-1}(x^\nu-a^\nu)(L^\mu_\nu +
    R^{-2} {1\over \sigma(a)+\sqrt{\sigma}} \eta_{\nu\rho} a^\rho
    a^\lambda L^\mu_\lambda )
\end{eqnarray}
where $\lambda=R^{-2}$ has been used. Eq.(\ref{A20})  is just
eq.(\ref{transformation}) in the text. } }

\end{appendix}

%\newpage


\begin{thebibliography}{99}
\bibitem{foot} In the ordenary non-relativistic quantum mechanics,
it is obvious that the  time everlution and space displasememts are
independent of dynamics. Namely, $\hat{H}=i\hbar \pa_t,\;
\hat{\pi}=-i\hbar \pa_x$ are always  true to any dynamics system
with $H=\pi^2/2m + V(x,t,\pi)$. As commutitator  between generators
of space-time displasements, $[\hat{H},\; \hat{\pi}]=\hbar^2
[\pa_t,\; \pa_x]=0.$
\end{thebibliography}
\end{document}